\documentclass[a4paper,11pt]{article}
\usepackage{jcappub} % for details on the use of the package, please
\usepackage[T1]{fontenc} % if needed

\newcommand{\beq}{\begin{equation}}
\newcommand{\eeq}{\end{equation}}

\newcommand{\be}{\begin{equation}}
\newcommand{\ee}{\end{equation}}

\newcommand{\bea}{\begin{eqnarray}}
\newcommand{\eea}{\end{eqnarray}}
\newcommand{\non}{\nonumber}

\title{Dark energy as a fixed point of the \\ Einstein Yang-Mills
 Higgs Equations }

\author[a,b]{Massimiliano Rinaldi}

\affiliation[a]{Dipartimento di Fisica, Universit\`a di Trento,\\  Via Sommarive 14, 38123 Trento, Italy. }
\affiliation[b]{INFN - TIFPA, \\  Via Sommarive 14, 38123 Trento, Italy.}
\emailAdd{massimiliano.rinaldi@unitn.it}

\abstract{ We study the  Einstein Yang-Mills Higgs equations in the $SO(3)$ representation on a isotropic and homogeneous flat Universe, in the presence of radiation and matter fluids. We map the equations of motion into an autonomous dynamical system of first-order differential equations and we find the equilibrium points. We show that there is only one stable fixed point that corresponds to an accelerated expanding Universe in the future. In the  past, instead, there is an  unstable fixed point that implies a stiff-matter domination. In between, we find three other unstable fixed points, corresponding, in chronological order, to radiation domination, to matter domination, and, finally, to a transition from decelerated expansion to accelerated expansion.  We solve the system numerically and we confirm that there are smooth trajectories that correctly describe the evolution of the Universe, from a remote past dominated by radiation to a remote future dominated by dark energy, passing through a matter-dominated phase.}
\begin{document}
\maketitle
 \flushbottom
%%%%%%

%%%%%%%%%%%%%%%%%
\section{Introduction}
%%%%%%%%%%%%%%%%%

\noindent The quest for a theoretically sound explanation for the current acceleration of the expansion of the  Universe is one of the major challenges of modern theoretical physics. Most proposed models can be roughly divided in two groups according to  whether they modify gravity or introduce some unusual matter content in the form of one or more scalar fields with appropriate potentials (for references see e.g. \cite{book,defelice}). In addition to these, there are interesting models based on non-local quantum field theory effects \cite{maggiore}.

The advances in observations have progressively constrained  the parameter space of dark energy models and, so far, $\Lambda$CDM is still the most favoured one \cite{PlanckXIV}. In this model, dark energy is sourced by some vacuum energy that appears as a constant term $\Lambda$ in the gravitational action and whose effect is, in the present epoch, contrasted by the cold dark matter component (CDM). The measured value of $\Lambda$ is several orders of magnitude smaller than any realistic estimation from quantum field theory (however, see \cite{roman} for  possible explanations) therefore it appears as a  very unnatural and fine-tuned explanation for the acceleration of the Universe. Future observations (such as the Euclid mission \cite{euclid}) should be hopefully able to reveal if the cosmological constant is truly a constant or it is a slowly varying field of some sort. 

In this paper we assume that dark energy is not a constant vacuum energy but it arises as a dynamical effect. Inspired by earlier work by Caldwell et al. \cite{spintessence}, and by the multifield dynamics studied in the context of Higgs inflation \cite{kaiser}, we  have  already explored in the past models of dark energy where gravity is implemented by the dynamics of gauge and Higgs fields \cite{max1,max2}. In particular, in \cite{max2} we have shown that an accelerated phase is possible when gravity is minimally coupled to a $SU(2)$ gauge fields and to a Higgs-like complex doublet.  

The Einstein Yang-Mills Higgs (EYMH) equations in a homogeneous and isotropic Universe were studied in few papers appeared before the discovery of the cosmic acceleration \cite{ochs,moniz}, so there was no motivation to find a source of dark energy. On the opposite, a lot of work has been done in inflationary models driven by gauge fields, eventually coupled to the Higgs field \cite{odintsov,Maleknejad:2011jw,Adshead:2012kp}. Finally, gravity non-minimally coupled to the Higgs field has been thoroughly investigated in the context of inflation \cite{shapo} and compact astrophysical objects \cite{namur}. 

In our previous work \cite{max2}, we did not take in account the coupling between the Higgs doublet and the gauge fields. In the present paper, we wish to include this coupling in the dynamical system to explore in full generality the physics of the EYMH action on a cosmological isotropic and homogeneous background. We also choose the representation $SO(3)$, instead of $SU(2)$ as we did in \cite{max2}, mainly to assess whether the late acceleration is some sort of gauge artefact or not. In fact, it is not, the acceleration is real and, as we will shortly see, it corresponds to the only stable fixed point of the system of equations.

One of the conceptual advantages of this model is that it  requires degrees of freedom whose dynamics is similar to the one of the standard model (SM). Therefore, there is no need to modify gravity or to introduce scalar fields with designer potentials that are quite unnatural in the realm of particle physics. Here,  the standard ``Mexican hat'' potential, analogous to the one of the SM Higgs field, is sufficient to lead to a final accelerated phase. However, as we will show, since there are no symmetry breaking effects, the shape of the potential can be relaxed to be some generic quartic potential. As already noted in \cite{max2}, the Mexican hat potential does not offer a ``slow-roll'' phase to the dynamics of the Higgs, it is simply way too steep to do that. However, there is another dynamical regime, dubbed ``ultra-slow roll'' in \cite{max2}, which is related to the kinetic energy of the Higgs phases and can explain the late-time domination of dark energy. In fact, the multifield dynamics of the Higgs components  modifies the Klein-Gordon equation in such a way that it prevents the Higgs field to reach its vacuum expectation value in a finite time. Thus, only in the infinite future the potential effectively vanish and the Higgs becomes constant. In the meanwhile, the potential behaves as a very slowly varying  effective cosmological constant equal to the square of the displacement of the Higgs field from its vacuum. In the present paper,  we will see that the same mechanism is at work in the $SO(3)$ representation, provided we keep account of the coupling between the Higgs triplet and the gauge fields.

The plan of the paper is the following. In the next section we display the action and the equations of motion that will be studied as a dynamical system in Sec.\ III. In Sec.\ IV we solve numerically the system for realistic initial conditions and we show that dark energy can be sourced by the dynamics of the EYMH equations. We conclude in Sec.\ V with some considerations.

%%%%%%%%%%%%%%%%%
\section{EYMH equations in FLRW spacetime}
%%%%%%%%%%%%%%%%%

\noindent Let us begin by considering the equations of motion obtained from the Lagrangian
\bea\label{lagra}
L=\sqrt{|\det g|}\left[  {M^{2}\over 2}R-{1\over 4}F^{a\mu\nu}F^{a}_{\mu\nu}-{1\over 2}(D_{\mu}\Phi^{a})(D^{\mu}\Phi^{a})-V(\Phi^{a}\Phi^{a})  \right]+L_{m}\,,
\eea
where 
\bea
F^{a}_{\mu\nu}=\partial_{\mu}A_{\nu}^{a}-\partial_{\nu}A_{\mu}^{a}+g\epsilon^{abc}A^{b}_{\mu}A^{c}_{\nu}\,,
\eea
and
\bea
D_{\mu}\Phi^{a}=\partial_{\mu}\Phi^{a}+g\epsilon^{abc}A^{b}_{\mu}\Phi^{c}\,,
\eea
is the covariant derivative with the coupling constant $g$. The term $L_{m}$ denotes the standard Lagrangian of radiation and matter fields in the form of perfect fluids. We choose the representation $SO(3)$ so latin indices run over $(1,2,3)$ and summation is understood.
As for the potential, we choose the standard ``Mexican hat'' profile
\bea\label{pot}
V={\lambda\over 4}\left(\Phi^{2}-\Phi_{0}^{2}\right)^{2}\,,
\eea
where, from now on, $\Phi^{2}\equiv\Phi^{a}\Phi^{a}$. The value of the vacuum term $\Phi_{0}^{2}$ it is not known but, as we will see below, this is not really relevant for our purposes.  We stress once more that  we are not identifying this potential with the SM potential, we just take it as a template for our investigations.

We assume that the metric of spacetime is isotropic and homogeneous, namely
\bea
ds^{2}=-N^{2}(t)dt^{2}+a^{2}(t)\delta_{ij}dx^{i}dx^{j}\,.
\eea
This implies that the symmetry of spacetime ``overrides'' the gauge symmetry, reducing the effective degrees of freedom of the gauge field to just one, according to \cite{galtsov,maroto}
\bea\label{wgauge}
A^{a}_{0}=0\,,\quad A_{i}^{a}=f(t)\delta^{a}_{i}\,.
\eea
This choice guarantees that  isotropy  and homogeneity of space-time are preserved. Similar considerations work for static spherically symmetric solutions of Yang-Mills theories \cite{bartnik,clement,volkov}.

With the constraint \eqref{wgauge} we  have
\bea
F^{a\mu\nu}F^{a}_{\mu\nu}=6\left({g^{2}f^{4}\over a^{4}}-{\dot f^{2}\over N^{2}a^{2}}\right)\,,
\eea
and 
\bea
(D_{\mu}\Phi^{a})(D^{\mu}\Phi^{a})=-{\dot\Phi^{2}\over N^{2}}+{2g^{2}f^{2}\Phi^{2}\over a^{2}}\,,
\eea
where the dot stands for a derivative with respect to $t$ and $\dot\Phi^{2}\equiv\dot\Phi^{a}\dot\Phi^{a}$

We now replace these expressions into the Lagrangian \eqref{lagra}, we  work out the equations of motion by variation of the fields $N$, $a$, $f$, and $\Phi^{a}$, and we set $N(t)= 1$ at the end. The Friedmann equations then read
\bea\label{Hsq}
H^{2}&=&{1\over 3M^{2}}\left[ {3\dot f^{2}\over 2a^{2}}+{3g^{2}f^{4}\over 2a^{4}}+{\dot\Phi^{2}\over 2} +{g^{2}f^{2}\Phi^{2}\over a^{2}}+V+\rho_{\rm m}+\rho_{\rm r} \right]\,,\\\label{Hdot}
\dot H&=&-{1\over 2M^{2}}\left[ {2\dot f^{2}\over a^{2}}+{2g^{2}f^{4}\over a^{4}}+\dot \Phi^{2}+{2g^{2}f^{2}\Phi^{2}\over 3a^{2}}+\rho_{\rm m}+{4\rho_{\rm r} \over 3} \right]\,,
\eea
where $H=d\ln a/dt$ and matter and radiation densities satisfy respectively the equations
\bea\label{eqmatt}
\dot\rho_{\rm m}=-3H\rho_{\rm m}\,,\quad \dot\rho_{\rm r}=-4H\rho_{\rm r}\,.
\eea
The system is implemented by the equation for the gauge field degree of freedom 
\bea\label{eqf}
\ddot f+H\dot f+{2g^{2}f^{3}\over a^{2}}+{2g^{2}f\Phi^{2}\over 3}=0\,,
\eea
and by the Klein-Gordon equations for each component of $\Phi^{a}$
\bea\label{eqkg}
\ddot \Phi^{a}+3H\dot\Phi^{a}+{2g^{2}f^{2}\Phi^{a}\over a^{2}}+\lambda\Phi^{a}(\Phi^{2}-\Phi_{0}^{2})=0\,,\quad a=1,2,3\,.
\eea
As anticipated in the introduction, and similarly to what happens in the case studied in \cite{max2},  $\Phi_{0}^{2}$ cannot be an exact solution of the last equation, unless $f=0$ or $g=0$ or $\Phi_{0}=0$. In principle, however, the condition $\Phi^{2}=\Phi_{0}^{2}$ can be achieved in the infinite future, if $a$ diverges but $f$ does not.  As already explained in ref.\ \cite{max2}, an eventual  late-time dynamics, dominated by a slowly varying effective cosmological constant,  cannot be described by the standard slow-roll condition (in fact $V^{-1}dV/d\Phi$ diverges in the large $N$-limit) but rather by a ``ultra slow-roll'' regime, where both $\ddot\Phi$ and $\dot\Phi$ can be neglected in eq.\ \eqref{eqkg}, while the last two terms take over and yields a secular variation of the potential.

This interpretation is further supported by noting that the right hand side of eq.\ \eqref{Hdot} is negative definite, hence an exact de Sitter-like solution, with $H=$ const, is not allowed. However, one can look for asymptotic solutions characterised by $H=$ const in the infinite future, with $a\rightarrow\infty$  and the Higgs field settled at the value $\Phi^{2}=\Phi_{0}^{2}$ so that its time derivative vanishes.

We now consolidate these qualitative discussions by studying the system of equations and its equilibrium points.

%%%%%%%%%%%%%%%%%
 \section{Dynamical system analysis}
 %%%%%%%%%%%%%%%%%

 \noindent A very efficient way to study systems of equations as the one in the previous section, especially in the context of dark energy, is to map it into a closed system of first-order differential equations by defining a new set of dimensionless variables \cite{book}. This technique is also ideal for numerical treatment of the equations. We choose the following variables  
  \bea\label{vars}
 x&=&{f'\over \sqrt{2}aM}\,,\quad y={gf^{2}\over \sqrt{2}MHa^{2}}\,,\quad v={1\over MH}\sqrt{V\over 3}\,,\quad r={1\over MH}\sqrt{\rho_{\rm r}\over 3}\,,\\\non
l&=&{\sqrt{2}Ma\over f}\,,\quad w^{i}={gf\Phi^{i}\over \sqrt{3}MaH}\,,\quad z^{i}={(\Phi^{i})'\over \sqrt{6}M}\,,\quad i=1,2,3\,,
 \eea
where we introduced the derivative with respect to the e-folding number $N=\ln a$, denoted by a prime. The conversion between time derivative and $N$-derivative is  $\dot X= HX'$ for any function of time $X$.

The  deceleration parameter, defined by  $q=-1-\dot H/H^{2}\equiv -1-H'/H$, indicates whether the expansion of the Universe is accelerated or not. When only one source of matter is present $q$ is related to the equation of state parameter $w$ by the equation $q=(1+3\omega)/2$: whenever $q<0$ ($\omega>-1/3$) the expansion is accelerated. In our case, the acceleration is a combined effect of the gauge and the Higgs fields, therefore define an effective equation of state parameter $\omega_{\rm eff}$ such that $q=(1+3\omega_{\rm eff})/2$. By using the Friedmann equations and the variables above we find that
 \bea\label{qdef}
 q=\frac12(1+x^{2}+y^{2}+r^{2}-3v^{2}+3z^{2}-w^{2})\,,
 \eea
 where we used the short-hand notation
 \bea
  z^{2}\equiv (z_{1})^{2}+(z_{2})^{2}+(z_{3})^{2}\,,\quad w^{2}\equiv  (w_{1})^{2}+(w_{2})^{2}+(w_{3})^{2}\,.
   \eea
With these expressions, eq.\ \eqref{Hsq} can be written as the constraint
\bea\label{omm}
\Omega_{\rm m}\equiv{\rho_{\rm m}\over 3M^{2}H^{2}}=1-(x^{2}+y^{2}+z^{2}+w^{2}+v^{2}+r^{2})\,,
\eea
in terms of the matter density $\Omega_{\rm m}$. Analogously, we can define the densities of radiation and dark energy respectively as
\bea\label{omrad}
\Omega_{\rm r}&=&r^{2}\,,\\\label{omde}
\Omega_{\rm de}&=&x^{2}+y^{2}+z^{2}+w^{2}+v^{2}\,,
\eea
so that 
\bea\label{omconst}
1=\Omega_{\rm de}+\Omega_{\rm r}+\Omega_{\rm m}\,.
\eea
By differentiating with respect to $N$ each of the variables \eqref{vars}, and by using the equations \eqref{eqmatt}-\eqref{eqkg}, we find a closed system of eleven first-order differential equations, given by
\bea\label{fullsys}
l'&=&l(1-lx)\,,\\\non
 x'&=&(q-1)x-l(w^{2}+2y^{2})\,,\\\non
 y'&=&y(q-1+2xl)\,,\\\non
 r'&=&(q-1)r\,,\\\non
v'&=&v(q+1)+\alpha l (w_{1}z_{1}+w_{2}z_{2}+w_{3}z_{3})\,,\\\non
w_{i}'&=&w_{i}(q+lx)+\sqrt{2}lyz_{i}\,,\quad i=1,2,3\,,\\\non
 z_{i}'&=&(q-2)z_{i}-lw_{i}(\sqrt{2}y+\alpha v)\,,\quad i=1,2,3\,,
\eea
where $\alpha={\sqrt{3\lambda}/ g}$ is a dimensionless constant whose value will be discussed below.

Before solving numerically the system, it is crucial to find the equilibrium points and to study their stability. The system has an infinite number of fixed points that, however, correspond to a discrete and finite set of values for the deceleration parameter, namely $q=2,1,1/2,0,-1$. We now examine each family of equilibria.
\vspace{0.5cm}

\noindent $\mathbf{q=2}$: in this case the set of fixed points is defined by
\bea\non
z^{2}=1\,, \quad x = 0\,,\quad y = 0\,,\quad v = 0\,,\quad r = 0\,,\quad l = 0\,,\quad w_{i} = 0\,,\quad i=1,2,3\,.
\eea
Physically, this solution corresponds to $f=0$ and $V=0$, which means that the Higgs field takes its vacuum value $\Phi^{2}=\Phi_{0}^{2}$ on the fixed point. By computing the Jacobian we find that the eigenvalues are all real but some of them are vanishing thus the equilibrium points are non-hyperbolic and numerical methods are necessary to fully assess the stability. In fact, the numerical solution (see below) shows that this is an unstable equilibrium that corresponds to a Universe dominated by stiff matter ($\omega_{\rm eff}=1$) in the remote past.
\vspace{0.5cm}

\noindent $\mathbf{q=1}$: in this case we have two classes of solutions, namely
\bea\non
(a)&& x^{2}=1-r^{2}\,,\quad l^{2} = (1-r^{2})^{-1}\,,\quad y = 0\,,\quad v = 0\,,\quad z_{i} = 0\,,\quad w_{i}=0\,,\quad i=1,2,3\,,\\\non
(b)&& r^{2}=1-x^{2}-y^{2}\,,\quad  v = 0\,,\quad  l = 0\,,\quad z_{i} = 0\,,\quad w_{i}=0\,,\quad i=1,2,3\,.\eea
Both classes are non-hyperbolic  since some eigenvalues are vanishing. For the class $(a)$ we see that the radiation domination cannot be realised entirely with relativistic matter, which would require $r=1$ and a diverging  $l$. For the class $(b)$ instead, $r=1$ is allowed provided $x=y=0$, therefore radiation domination can be achieved with relativistic matter only. As for the previous case, $v=0$ implies that  $\Phi^{2}=\Phi_{0}^{2}$. The numerical solution shows that this equilibrium is not stable, in fact it is ``metastable'' because the condition $q= 1$ (i.e. a radiation-dominated Universe with $\omega_{\rm eff}=1/3)$ can last for few e-foldings, for realistic initial conditions.

\vspace{0.5cm}

\noindent $\mathbf{q=1/2}$: here we find the first hyperbolic equilibrium point, which is located at the origin, where all variables vanish. The eigenvalues are all real and have both positive and negative sign, thus the equilibrium is unstable. Physically, this solution corresponds to a Universe dominated by matter ($\omega_{\rm eff}=0$). By solving the system numerically with the initial conditions set arbitrarily close to the equilibrium point, we find that the matter content decreases in few e-foldings while dark energy gradually takes over. The radiation content instead does not increase since its linearised equation yields $r\sim e^{-N/2}$. Similarly to the previous case, this fixed point can represent the metastable matter domination of the Universe, which ends when dark energy becomes dominant.

\vspace{0.5cm}

\noindent $\mathbf{q=0}$: we have another non-hyperbolic fixed point located at
\bea\non
w^{2}=1\,,\quad x = 0\,,\quad y = 0\,,\quad v = 0\,,\quad r = 0\,,\quad l = 0\,,\quad  z_{i}=0\,,\quad i=1,2,3\,.
\eea
The numerical analysis reveals that it is an unstable point and, in the global evolution, it marks the transition between deceleration ($\omega_{\rm eff}>-1/3$) and acceleration ($\omega_{\rm eff}<-1/3$). Note that  $q$ vanishes because of the negative sign of  $w^{2}$  in the definition \eqref{qdef}. In typical quintessence models, the field $\Phi$ does not appear in the equations of motion, only $\dot\Phi$ does. In our model, instead,  the Friedmann equations depend also on $\Phi$ because of the coupling  to the gauge degree of freedom $f$ via the constant $g$. Therefore we are forced to define the variables $w_{i}$ (usually absent in quintessence models), which have a non-trivial dynamics and allow for the appearance of the critical point at $q=0$. In other words, in our model the takeover of dark energy (or more precisely, the transition to the accelerated phase) crucially depends on the coupling between the Higgs and the gauge field.

\vspace{0.5cm}

\noindent $\mathbf{q=-1}$: this is the most relevant equilibrium point as it corresponds to an accelerated Universe ($\omega_{\rm eff}=-1$). It is hyperbolic and  located at  \footnote{Strictly speaking, there are two fixed points with $v=\pm 1$ but, physically, $v$ is definite positive for an expanding Universe.}
\bea
x = 0\,,\quad  y = 0\,,\quad  v = 1\,,\quad  r = 0\,,\quad  l = 0\,,\quad  z_{i} = 0\,,\quad w_{i} = 0\,,\quad i=1,2,3\,.
\eea
All eigenvalues are real and negative except for one so the point is a saddle point. By linearising the system, we discover that  it is the variable $l(N)$ that runs away from the fixed point. In fact,  by solving the linearised equation for $y$, we find that the scale factor evolves as $a/f\sim e^{N}$. In addition, $\Phi_{a}'\sim e^{-3N}$, which indicates that the Higgs field components tend to constant values. This is consistent since $v\rightarrow1$ nearby the fixed point, so that 
\bea
(\Phi^{2}_{\infty}-\Phi_{0}^{2})= 2\sqrt{3\over\lambda}MH_{\infty}\,.
\eea 
Finally, the radiation energy density exponentially decays to zero as it can be easily seen from the linearised equation for $r$. Physically, all this means that the Higgs field tends to settle asymptotically on some value (which does not necessarily coicide with the vacuum expectation value $\Phi_{0}$), dragging the Universe towards an accelerated phase, where the Yang-Mills field is constant, radiation and matter fade away, while the potential becomes an effective cosmological constant. This final state corresponds to the only stable equilibrium point therefore we conclude that the EYMH system for a homogeneous and flat Universe has a final dark energy-dominated phase. 
\vspace{0.5cm}

Before proceeding into the discussion of the numerical solutions, we must assign a reasonable value to $\alpha$, which is in fact the only free parameter of the theory. We note that the stability of the equilibrium points is independent on this parameter since the eigenvalues do not depend on $\alpha$. This can be qualitatively understood from the fact that in the system \eqref{fullsys} $\alpha$ appears only in non-linear terms. Therefore, the structure of the fixed points in unaffected by the value of $\alpha$. However, for numerical solutions we need to estimate it.

In the Higgs sector of the SM, the coupling constants are related to the vacuum energy and to the mass spectrum of the theory. In particular, $g=2M_{W}/v$, where $M_{W}\simeq 80$ GeV is the mass of the $W^{\pm}$ bosons and $v=(\sqrt{2}G_{F})^{-1/2}\simeq 246$ GeV is the vacuum expectation value of the Higgs, determined by the Fermi coupling constant. The self-coupling of the Higgs is determined experimentally to be  $\lambda=0.13$. If we identify the parameters $g$ and $\lambda$ in our model with these values, we find $\alpha=0.96$ \footnote{It is amusing how close this value is to the inflationary scalar spectral index $n_{s}$.}.
With no other available guiding principle, we then fix $\alpha=1$ in our numerical calculations and we will discuss the physical observables at the end.

%%%%%%%%%%%%%%%%%
\section{Numerical solutions}
%%%%%%%%%%%%%%%%%

\noindent In the previous section we have shown the existence of one stable asymptotic equilibrium point that corresponds to a final state of dark energy domination. In this section we solve the dynamical system  numerically in the attempt to build a realistic scenario. We anticipate that the numerical solutions shown below are just a preliminary study of the full solution space, whose exploration requires sophisticated numerical tools and will be presented in another work. Our aim is just to demonstrate the validity of our model in recreating a realistic scenario.

We choose to fix the initial conditions at the present time, by imposing $\Omega_{\rm r}\approx 0$, $\Omega_{\rm de}\approx 0.69$, and $\Omega_{\rm m}\approx 0.31$.
In detail, we set
\bea\label{incond}
\left.\begin{array}{llll} x = 10^{-9}\,, & y = 10^{-8}\,, & v = 0.83 \,,& r = 10^{-2}\,, \\ z_{1} = -10^{-11} \,,& z_{2} =2\times 10^{-11}\,, & z_{3} = 3\times 10^{-11} \,, & l = 4\times 10^{-8} \,,\\w_{1} =10^{-9} \,,& w_{2} = 4\times 10^{-10}\,, & w_{3} = 2\times 10^{-9} \,,& \end{array}\right.
\eea
at $N=0$.  Our numerical solutions show that, with these initial data, matter-dark energy equality occurs at $N\simeq-0.3$ (corresponding to a redshift ${\rm z}\simeq 0.3$) while matter radiation equality is at $N\simeq -8.1$ (redshift ${\rm z}=3360$).
 
\begin{figure}[ht]
  \centering 
  \includegraphics[scale=0.4]{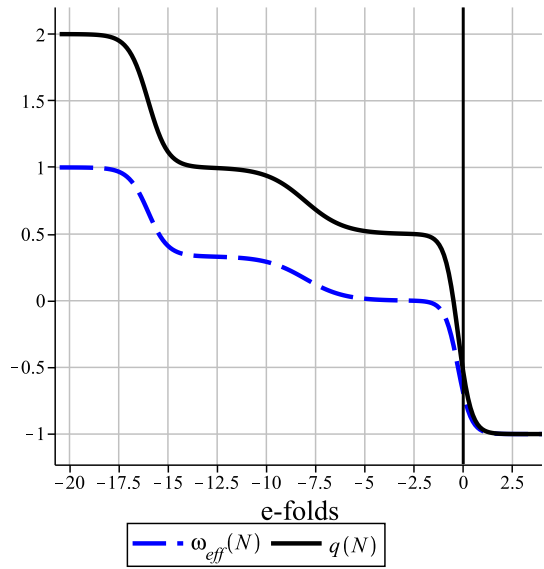} 
   \caption{Evolution of the deceleration parameter $q(N)$ and the corresponding effective equation of state $\omega_{\rm eff}(N)$. The initial conditions are given by \eqref{incond} at $N=0$, corresponding to the black vertical line. We note that the evolution of the system mimics the transition between a remote past stiff matter-dominated Universe and a final dark energy-dominated Universe, passing through two transient phases of radiation and matter domination respectively. }
  \label{Qglobal}
  \end{figure}

The first relevant plot is in Fig.\  (\ref{Qglobal}) that shows the behaviours of the deceleration parameter and of $\omega_{\rm eff}$ as a function of $N$ over several e-foldings. We see that the remote Universe starts with a stiff matter-dominated phase ($\omega_{\rm eff}=1$) and evolves towards the equilibrium point at $q=-1$, which corresponds to $\omega_{\rm eff}=-1$. The transition between the two extrema is characterised by two plateaux corresponding to radiation and matter domination respectively. They also coincide with two unstable fixed points ($q=1$ and $q=1/2$) discussed in the previous section. This proves that the Yang-Mills Higgs equations coupled to gravity and ordinary fluids are able to describe all phases of the evolutions of the Universe (except for the initial inflationary expansion). Initial conditions may change the duration of each intermediate phase but, in general, do not modify the overall evolution of the solutions.

\begin{figure}[ht]
  \centering 
  \includegraphics[scale=0.4]{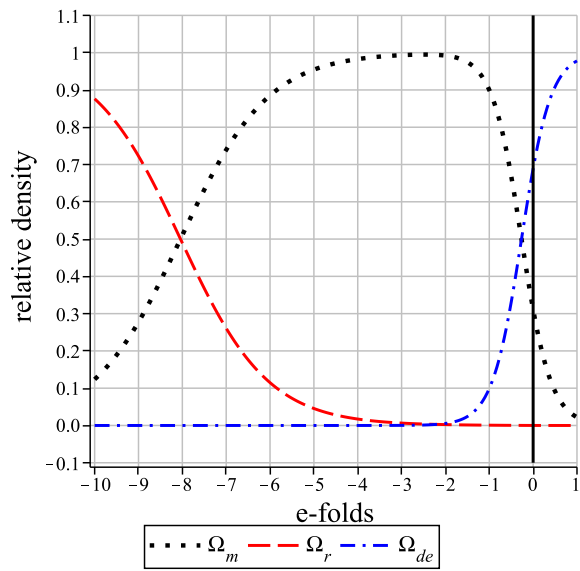} 
   \caption{Evolution of the density parameters with the same initial conditions as in Fig.\ (\ref{Qglobal}). The vertical black line is the present time.}
  \label{OmegaGlobal}
  \end{figure}

\begin{figure}[ht]
  \centering 
  \includegraphics[scale=0.4]{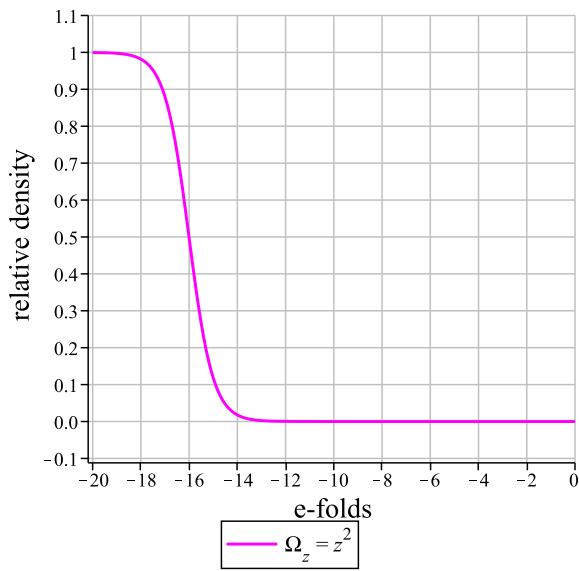} 
   \caption{Evolution of the ``stiff matter'' density parameter $\Omega_{\rm z}=z_{1}^{2}+z_{2}^{2}+z_{3}^{2}$, which is dominant in the remote past. We find that the time at which it becomes dominant only depends on the initial values given to $z_{i}$ at $N=0$. By lowering these values we can push back the stiff-matter phase at virtually arbitrarily large time $|N|$. While mathematically interesting, this early time solution is not compatible with observations, as it does not connect smoothly to an inflationary phase.}
  \label{OmegaZ}
  \end{figure}

\begin{figure}[ht]
  \centering 
  \includegraphics[scale=0.4]{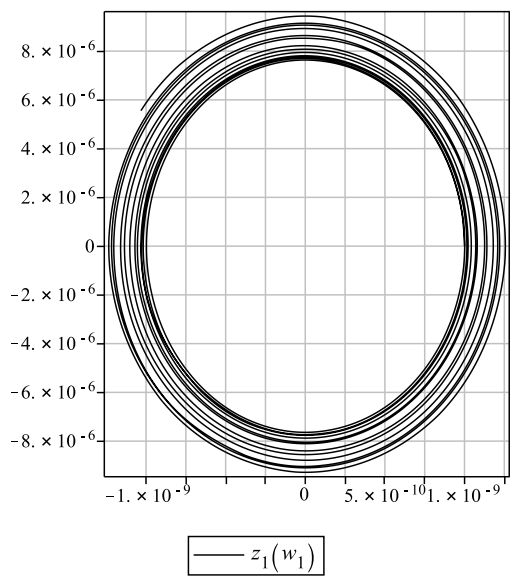} 
   \caption{Evolution of the variable $z_{1}$ as a function of $w_{1}$ in the range $N=[-0.03,0]$.}
  \label{zofw}
  \end{figure}

The evolution of the relative density of radiation, matter, and dark energy is displayed in Fig.\ (\ref{OmegaGlobal}). If we extend the computation at earlier times, we find that the stiff matter domination is caused by the variable $z^{2}$, which becomes dominant in the remote past, see Fig.\ (\ref{OmegaZ}). It is also interesting to see the behaviour of other variables. In particular, we show in Fig.\ (\ref{zofw}) the plot of $z_{1}$ as a function $w_{1}$ in the present epoch (the results for the other two couples of variables are qualitatively the same). The plot clearly shows the attractive nature of the fixed point with $q=-1$.

As explained above, in principle we do not have clear indications on the value of $\alpha$. However, we can relate it to physical observables. By using the definitions of the variables $l(N)$, $y(N)$, and $v(N)$, we find that
\bea
{M\over H}&=&{yl^{2}\over \sqrt{2}g}\,,\\\non
 {\Phi^{2}-\Phi_{0}^{2} \over M^{2}}&=&{6\sqrt{2}v\over \alpha y l^{2}}\,.
\eea
The first relation implies that, at the present time, and with the initial conditions \eqref{incond}, the coupling $g$ must be very small. With $H=1.4\times 10^{-42}$ GeV and $M=2.4\times 10^{18}$ Gev,  and $\alpha=1$, we find $g\sim 10^{-54}$. For $\alpha=1$, this implies an even smaller $\lambda$, of the order $\lambda \sim 10^{-109}$. The second relation instead gives an estimate of the  displacement from the vacuum value of the Higgs field, which corresponds to $(\Phi^{2}-\Phi_{0}^{2})/M^{2}\sim 10^{-6}$ with  \eqref{incond}.

If we wish to increase the value of $g$ and $\lambda$, we need to increase of several orders of magnitude the initial value of $l(N)$ (on the contrary, the value of $y(N)$ cannot be larger than unity because of the constraints \eqref{omrad} and \eqref{omconst}). In turn, this implies that the initial value of $x(N)$ must be carefully chosen so that the first equation of \eqref{fullsys} does not yield large derivatives. This means, essentially, that $x\approx 1/l$ over all the integration range. All these refinements require more powerful numerical integration methods and will be discussed elsewhere. However, the solutions shown above are promising and we believe that a thorough analysis will be able to find stable numerical solutions with larger values of $\lambda$ and $g$.

%%%%%%%%%%%%%%%%%
\section{Concluding remarks}
%%%%%%%%%%%%%%%%%

\noindent In this paper, we have studied the full Einstein Yang-Mills Higgs system of equation embedded in a flat Robertson-Walker spacetime. The remarkable feature of this system is the set of fixed points that coincide with various kinds of matter domination, beginning with  an unstable stiff matter type in the remote past and ending with a stable final stage of dark energy domination. The transition between these two eras is characterised by metastable phases of radiation and matter domination. Numerical solutions seem to confirm this evolution and we conclude that this model is valid explanation of the current acceleration of the Universe. 

Of course there are many issues to be considered. First of all, a full numerical study of the system is necessary, in particular to assess the stability against perturbations. In addition, the model can be extended to different kinds of gauge groups and/or potentials. In any case, our task was to show that EYMH equations were able to reproduce dark energy. We found that not only this is true but also that the entire evolution from radiation domination to today is compatible with this model.

It would be interesting to investigate extensions of this model towards the inflationary era. This might be in principle possible recalling that the non-minimal coupling of the Higgs field or of the gauge potential to gravity provides for valid models  of inflation \cite{Maleknejad:2011jw,Adshead:2012kp,shapo}. Typically, the non-minimal coupling becomes negligible at low energy so it would not affect the evolution of the Universe after inflation, which is well described by our model.  The possibility that the  EYMH Lagrangian  \emph{non-minimally coupled to gravity}  connects the current accelerate expansion to the initial inflationary phase is therefore an open and interesting question.

\noindent {\bf Acknowledgment} The authors would like to thank S.\ Zerbini, L.\ Vanzo, and G.\ Cognola (U.\ Trento)  and A.\ F\"uzfa (U.\ Namur) for helpful discussions.

%%%%%%%%%%%%%%%%%%%%%%%%%%%%%%

%%%%%%%%%%%%%%%%%%%%%%%%%%%%%%

\end{document}